\documentclass[10pt, conf, twocolumn]{IEEEtran}
\usepackage[left=0.75in,right=0.75in,top=0.75in,bottom=0.75in]{geometry}

\usepackage{color}
\usepackage{tikz}
\usetikzlibrary{positioning}
\usetikzlibrary{shapes,snakes}
\usetikzlibrary{arrows,patterns}
\usepackage{qtree}
\usepackage{times}
\usepackage{epsfig}
\usepackage{graphicx}
\usepackage{epstopdf}
\usepackage{algpseudocode}
\usepackage{algorithm}
\usepackage{amsmath}
\usepackage{amssymb}
\usepackage{amsxtra}
\usepackage{multirow}
\usepackage{mathtools}
\usepackage{cleveref}
\usepackage{url}
\usepackage{cite}
\usepackage{amsthm}
\usepackage{adjustbox}
\usepackage[latin1]{inputenc}
\usepackage{caption}
\usepackage[caption = false]{subfig}

\newtheorem{theorem}{Theorem}
\newtheorem{lemma}{Lemma}

\newtheorem{corollary}{Corollary}
\newtheorem{assumption}{Assumption}
\newtheorem{remark}{Remark}
\newtheorem*{proof*}{Proof}

\algnewcommand{\Initialize}[1]{%
  \State \textbf{Initialize:}
  \Statex \hspace*{\algorithmicindent}\parbox[t]{.8\linewidth}{\raggedright #1}
}

\begin{document}

\title{Dynamic Spatio-Temporal Resource Provisioning for On-Demand Urban Services in Smart Cities}

\author{ \IEEEauthorblockN{\large Muhammad Junaid Farooq and Quanyan Zhu } \\ \IEEEauthorblockA{Department of Electrical \& Computer Engineering, Tandon School of Engineering, \\New York University, Brooklyn, NY 11201, USA,} Emails: \{mjf514, qz494\}@nyu.edu. \vspace{-0.1in}

}

\maketitle

\begin{abstract}
Efficient allocation of finite resources is a crucial problem in a wide variety of on-demand smart city applications. Service requests often appear randomly over time and space with varying intensity.
Resource provisioning decisions need to be made strategically in real-time, particularly when there is incomplete information about the time, location, and intensity of future requests. In this paper, we develop a systematic approach to the dynamic resource provisioning problem at a centralized source node to spatio-temporal service requests.
The spatial statistics are combined with dynamically optimal decision-making to derive recursive threshold based allocation policies. The developed results are easy to compute and implement in real-time applications. For illustrative purposes, we present examples of commonly used utility functions, based on the power law decay and exponential decay coupled with exponentially, and uniformly distributed intensity of stochastic arrivals to demonstrate the efficacy of the developed framework. Semi-closed form expressions along with recursive computational procedure has been provided. Simulation results demonstrate the effectiveness of the proposed policies in comparison with less strategic methodologies.  \vspace{-0.0in}
\end{abstract}


\IEEEpeerreviewmaketitle


\vspace{-0.15in}
\section{Introduction}
The integrated networks of engineered cyber and physical systems, referred to as the Internet of Things (IoT), provides the enabling technology for cities to greatly improve the security, life, and wellbeing of its citizens~\cite{socially_responsible}.
The smart cities paradigm is creating a plethora of opportunities to  efficiently utilize available city resources\footnote{The term `resources' may generically refer to emergency response units, taxis, wireless channels, aerial vehicles, vaccines, etc.}~\cite{city_of_things}.
Resource allocation problems occur in a wide variety of scenarios in smart cities such as in disaster management, emergency response systems, public safety systems, taxi pickups, controlling epidemic outbreaks, data collection using wireless sensors, etc.~\cite{resource_provisioning}.
Typically, there is a centralized \emph{source node} having a finite number of available resources that need to be allocated to demand nodes or \emph{service requests} that arrive sequentially over time at random locations with varying severity or \emph{intensity}.
The task of the source node is to allocate available resources to service requests in real-time to maximize the total expected utility\footnote{The term `utility' used in the paper may refer to different quantities such as social welfare, revenue, etc., according to the application scenario.} obtained from allocation.

\begin{figure}[t!]
	\centering
	\vspace{0.15in}
	\includegraphics[width=3.4in]{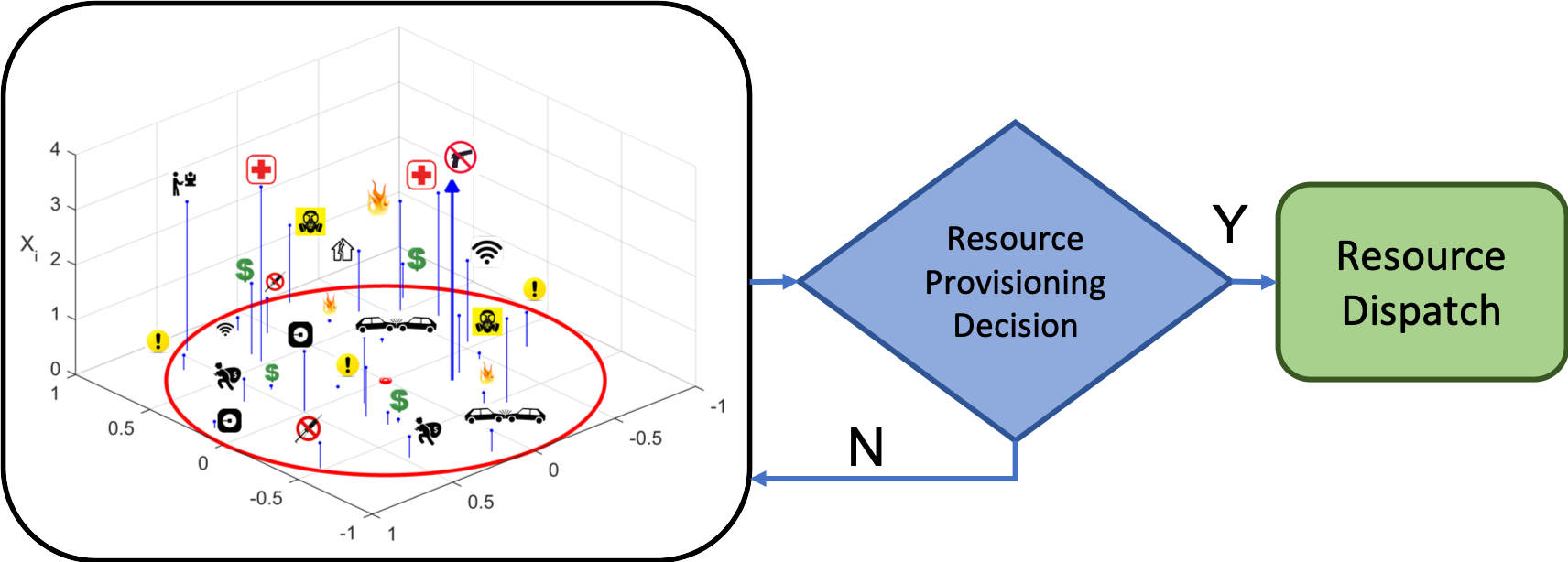}\\
	\caption{Illustration of the centralized resource allocation problem during one time slot. The impulse height represents the intensity of requests. The maximum intensity request is highlighted by a bold blue impulse with an arrowhead. \vspace{-0.1in}}\label{Fig:Sys_model}
\end{figure}



Fig.~\ref{Fig:Sys_model} illustrates a snapshot of the spatially disperse service requests that have accumulated over time with reference to a centralized source node placed at the origin serving a circular region. The height of impulses at the request locations represents the intensity or magnitude of the demand.
One of the potential ways to provision resources is to immediately allocate upon the arrival of a request regardless of its intensity or distance. However, it may result in myopic decisions leading to a suboptimal outcome. In many practical applications, the decision is made once a pool of requests is available. The challenge in the allocation decision by the source is two-fold. Firstly, a very high intensity demand request may need service but it may be located at a farther distance from the source. Secondly, the source needs to decide whether to allocate a resource to one of the current requests or to wait for future requests, which may result in higher benefit. While discarding allocation to current requests, the source might keep waiting for better requests that never arrive in the future. In essence, the question is where to draw the line for allocation in terms of distance and intensity of request as well as the waiting time of allocation.

As an example, consider a data collection problem where a centralized base station (BS) schedules the uplink transmission of spatially deployed sensor nodes. The sensor nodes randomly request for data collection to the BS informing it about their location as well as the channel state information\footnote{We assume that sensor nodes are always able to communicate low bandwidth signaling information to the BS via separate control channels.}. The base station then decides whether to collect data from the sensor or to discard it. If the sensor is close to the BS but the channel gain is extremely low, then the probability of successfully obtaining data from it might be low. However, if the sensor is located far away but the channel gain is extremely high, there might be a strong chance of successfully retrieving data from it.
Therefore, the BS needs to optimally select the nodes, from which to collect data, considering the limited number of time slots that are available to obtain data. Note that both the spatial and temporal components are highly important in the decision making while allocating decisions.






\vspace{-0.1in}
\subsection{Related Work}
Traditionally, resource allocation problems have been well studied in the Operations Research literature and are referred to as assignment problems. The classical bipartite assignment problem can be solved using the framework of Optimal Transportation~\cite{Villani_OT}, which is efficiently computed using Linear Programming for certain utility functions. Similarly, methods to solve the combinatorial optimization problems entailing to the assignment problem are available such as the Hungarian algorithm~\cite{hungarian}. Modified versions of these methods involving convex optimization have been developed to solve resource allocation problems in several applications, e.g., radio resource management in wireless communications~\cite{junaid_hungarian}, taxi dispatch~\cite{taxi_dispatch}, data collection~\cite{data_collection_magazine}, etc.
However, these techniques only present static and off-line solutions that cannot be implemented in real-time when there is incomplete information about future arrivals. Those that consider stochastic arrivals do not take the spatial aspect into account~\cite{sequential_stochastic}.




Recent works that attempt to tackle spatio-temporal resource provisioning problems in smart city applications include mobile data upload planning~\cite{data_collection}, cloud/fog computing resource allocation to applications~\cite{resource_provisioning}, communication resource management for network operators~\cite{network_resource_management}, urban sensing~\cite{urban_sensing}, and socially responsible resource distribution in smart cities~\cite{socially_responsible}. However, these works mainly use static assignment and are based on heuristic approaches to optimize the resource provisioning problem. Those that use arrival dynamics, e.g., in a crowd sensing application~\cite{data_collection_queueing}, make use of queueing models, in which the unserviced requests remain in the system. The realtime allocation of resources to stochastic incoming requests has been developed in~\cite{junaid_acc}. However, there is no waiting time involved and an allocation decision is made immediately upon arrival of requests. In practical applications, the decision making process may not be immediate. Hence, the source node may wait for an allocation period and decide to allocate a resource that may lead to the maximum benefit. In general, there is a lack of systematic and provably optimal approaches for centralized resource allocation to spatio-temporal service requests.




\vspace{-0.1in}
\subsection{Contributions}
In this work, we attempt to combine the spatio-temporal aspect along with incomplete information about resource requests to devise an integrated resource provisioning framework.
We leverage ideas from the seminal work on stochastic assignment of sequentially arriving tasks to workers~\cite{sequential_stochastic}. However, the model has been enriched to encompass a more generic utility function that also incorporates the spatial dimension of the sequentially arriving requests.
We make use of spatial point processes in conjunction with order statistics and dynamic programming to develop an integrated and holistic policy framework that can act as the foundation for allocation and pricing in a wide variety of applications in the context of smart city applications.

The rest of the paper is organized as follows: Section~\ref{Sec:Sys_model} presents an elaboration of the system model and problem description, Section~\ref{Sec:Mechanism} provides details on the solution methodology, Section~\ref{Sec:Results} contains numerical and simulation results, while Section~\ref{Sec:Conclusion} concludes the paper with insights on future research directions.

\vspace{-0.1in}
\section{System Model} \label{Sec:Sys_model}
Consider a typical source node at the origin with a discrete number of identical resources denoted by $N \in \mathbb{Z}^+$. The source is assumed to have an omni-directional service range of $R \in \mathbb{R}^+$. Service requests dynamically appear inside the served region according to a spatio-temporal Poisson Process with intensity $\lambda(r,\theta,t): [0,R] \times [0,2 \pi] \times \mathbb{R}^+ \rightarrow \mathbb{R}^+$, where the pair ($r$,$\theta$) represent the Polar coordinates of the service requests. Each service request is characterized by the tuple ($X_i,D_i$), where $X_i \in \mathbb{R}^+$ denotes the intensity of the request and $D_i \in [0,R]$ denotes its distance from the source node. It is assumed that both the distance and the intensity of requests is known at the source. Further, we assume that once a resource is allocated to service request, it becomes unavailable for allocation in the future.

\vspace{-0.1in}
\subsection{Characterization of Service Requests}
We assume a time slotted system with $\mathbf{t} = [1, 2, \ldots, T-1, T]$, where each time slot of duration $\tau$ represents an allocation period, also referred to as the decision horizon. To avoid non-uniformity in the average number of requests within allocation periods, we assume that the requests are uniform in the temporal domain, i.e., $\lambda(r,\theta,t)= \tau \tilde{\lambda}(r,\theta)$, where $\tau \in \mathbb{R}^+$ is a constant. Note that this assumption is not restrictive, since otherwise the width of time slots can be adjusted to create a uniform temporal profile of requests. In each time slot, the service requests are distributed spatially according to a Poisson Point Process (PPP). It implies that the number of requests in a circle of radius $R$ during one allocation period, denoted by $K$ follows a Poisson distribution with average density $\Lambda$. The probability density function (pdf) of $K$ can be expressed as follows:
\begin{align}
\mathbb{P}(K = k) = e^{-\Lambda } \frac{(\Lambda)^k}{k!},
\end{align}
where the, average density of service requests during each allocation period, can be evaluated as follows:
\begin{align}
\Lambda = \mathbb{E}[K] =  \int_{0}^{2 \pi} \int_{0}^{R} \tau \tilde{\lambda}(r,\theta) r {\rm d} r {\rm d} \theta.
\end{align}

Each service request has a type that represents the severity, criticality, or magnitude of demand. We model it as independent and identically distributed (i.i.d.) random variables $X_i$, $i \in \{1,\ldots,K\}$, for each allocation period. Furthermore, we assume that the pdf, denoted by $f_X(x)$, and cumulative distribution function (cdf), denoted by $F_X(x)$, of the intensity is known at the source\footnote{In practical situations, the statistical information about service requests can be obtained using spatio-temporal estimation techniques~\cite{spatiotemporal_estimation}.}. In order to exclude trivial cases in the resource allocation problem, we exclude the possibility of having no service request during an allocation period. It is stated formally stated as follows.
\begin{assumption}
We assume that there is at least one service request in every allocation period. Therefore we use the zero-truncated Poisson distribution to characterize the pdf of the number of service requests as follows:
\begin{align}
\mathbb{P}(K = k| K > 0 ) =  \frac{e^{-\Lambda }(\Lambda )^k}{(1 - e^{- \Lambda })k!}.
\end{align}
\end{assumption}

From the perspective of the source node, the distance of each service requests during an allocation period is also a random variable, which is independent of the intensity. The probability distribution of the distances can be expressed by the following lemma.
\begin{lemma}\label{dist_pdf}
The pdf of the distance $D$, of a randomly selected service request inside a circular region of radius $R$ from the source node, can be expressed as follows:
\begin{align}
f_{D}(d) = \frac{\int_0^{2 \pi} d \tilde{\lambda}(d,\theta)   {\rm d} \theta}{\int_0^{2 \pi} \int_0^R \tilde{\lambda}(r,\theta) r  {\rm d} r {\rm d} \theta}.
\end{align}
\begin{proof}
See \textbf{Appendix~\ref{dist_pdf_proof}}.
\end{proof}
\end{lemma}

\vspace{-0.1in}
\subsection{Utility of Resource Allocation}
We assume a generic utility function that characterizes the benefit obtained by allocating a resource to a service request of intensity $X_i$ that is located at a distance of $D_i$ from the source. The utility is denoted by $U(X_i,D_i): \mathbb{R}^+ \times [0,R] \rightarrow \mathbb{R}^+$. The utility function is assumed to be monotonically increasing in $X_i$ and monotonically decreasing in $D_i$\footnote{It implies that the utility is higher for close and high intensity requests and vice versa.}, since allocating a resource at a higher distance may incur additional cost. If the utility of obtained by allocating a resource to each service request during a single time slot is denoted by $Z_i = U(X_i,D_i)$, then $Z_i$ is also a random variable with the i.i.d. property formally expressed in the following remark.
\begin{remark}
The random variables $Z_i = U(X_i, D_i), \ i =1,\ldots,K$, are i.i.d. random variables since $X_i$ and $D_i$ are i.i.d., respectively.
\end{remark}

For notational convenience, we will henceforth drop the index $i$ and refer to the random variables describing the utility as $Z$. The cdf and pdf of the random variable $Z$ can be evaluated as follows:

\begin{lemma}
The cdf and pdf of the random variable $Z$ describing the utility of allocation to a randomly selected service request can be evaluated, respectively as follows:
\begin{align}
F_Z(z) = \mathbb{P}[Z \leq z] = \underset{\mathcal{S}}{\int \int} f_X(x) f_{D}(d) {\rm d} x {\rm d} d,
\end{align}
and
\begin{align}
f_Z(z) = \frac{d}{dz} \underset{\mathcal{S}}{\int \int} f_X(x) f_{D}(d) {\rm d} x {\rm d} d,
\end{align}
where $\mathcal{S} = \{(x,d): U(x,d) \leq z\}$.
\end{lemma}


We define the maximum utility during each time slot $j \in \{1, \ldots, T\}$ as follows:
\begin{align} \label{maximal_utility}
\tilde{Z}_j = \underset{1 \leq i \leq K_j}{\max}  \ Z_i,
\end{align}
where $K_j, j =1,\ldots,T$ are i.i.d. Poisson random variables with mean $\Lambda$. Since the random variables $\tilde{Z}_j, j \in \{1,\ldots,T\}$, for all time slots are also i.i.d, we hereby drop the subscript and refer to it as $\tilde{Z}$.
Then using extreme value theory, the pdf of $\tilde{Z}$ can be expressed by the following lemma:
\begin{lemma}\label{max_pdf_lemma}
The pdf of $\tilde{Z} = \max \{ Z_1, Z_2, \ldots, Z_K \}$, where $\{Z_i\}_{1 \leq i \leq K}$ are i.i.d. random variables with cdf $F_Z(z)$ and pdf $f_Z(z)$, and $K$ is a Poisson random variable with mean $\Lambda$, can be expressed as follows:
\begin{align}
f_{\tilde{Z}}(z) = \frac{\Lambda  f_Z(z) e^{\Lambda  (F_Z(z) - 1)}}{1 - e^{-\Lambda }}.
\end{align}
\begin{proof}
See \textbf{Appendix~\ref{max_pdf_lemma_proof}}.
\end{proof}
\end{lemma}


\vspace{-0.1in}
\subsection{Problem Definition}
The source has a total of $T$ allocation periods during which it needs to allocate all the $N \leq T$ available resources to incoming requests.
Instead of delaying allocation decision until all the requests have appeared, the goal is to decide allocation in real-time. Therefore, a mechanism is required to allocate resources dynamically while maximizing the total expected utility obtained from allocation.
During each allocation period, the decision problem is whether to allocated a resource to one of the current utility maximizing requests or to wait for the next batch of requests to decide.

The resource provisioning problem can be formally expressed as follows:
\begin{align} \label{optimization_problem}
\underset{\{i_1, i_2, \ldots, i_T\} \in \xi}{\max} \sum_{j = 1}^{T} \mathbb{E}[\tilde{Z}_j q_{i_j}]
\end{align}
where $\xi$ is the set of all possible permutations of the integers $1,\ldots,T$ and the vector $\boldsymbol{q} = [q_1, q_2, \ldots, q_T]$ is such that $q_1 = q_2 = \ldots = q_N = 1$ and $q_{N+1} = q_{N+2} = \ldots = q_T = 0$.
The objective of the optimization is to find the permutation vector $[i_1, i_2, \ldots, i_T]$ that maximizes the total expected utility.


\vspace{-0.1in}
\section{Optimal Dynamic Allocation Mechanism} \label{Sec:Mechanism}
In this section, we describe the process to solve the optimization problem expressed in~\eqref{optimization_problem}.
However, we would first like to emphasize the fact that the decision problem is only relevant if $N \leq T$, i.e., the number of available resources are less than the number of allocation periods. In other words, if the number of resources is more than the number of allocation periods, then the optimal policy would be to allocate a resource to the utility maximizing request in every allocation period. A grid representing the possible pairs of $(T,N)$ is illustrated in Fig.~\ref{Fig:grid}.

\vspace{-0.1in}
\subsection{Optimization}
We denote the optimal value obtained by solving the optimization problem in~\eqref{optimization_problem} if $T$ allocation periods are remaining and $N$ resources are available as $V(T,N)$. Let $\rho_T^N \in \mathbb{R}^+$ denote the  decision threshold on the random variable $\tilde{Z}$ if $T$ allocation periods are remaining and $N$ resources are available. Then, the value function can be expressed recursively using the following lemma:
\begin{lemma}\label{Value_lemma}
	If $T$ allocation periods are remaining and $N$ resources are available, then the total value obtained, in the case when a resource is allocated to the request and in the case when the request is discarded can be expressed as follows:
	\begin{align}
	V(T,N) =
	\left\{
	\begin{array}{ll}
	\tilde{Z}_T  + \mathbb{E}[V(T-1,N-1)],& \mbox{if  } \ \tilde{Z} \geq \rho_T^n, \\
	\mathbb{E}[V(T-1,N)], & \mbox{if  } \ \tilde{Z} < \rho_T^n.
	\end{array}
	\right.
	\end{align}
	\begin{proof}
		See \textbf{Appendix~\ref{Proof_value_lemma}}.
	\end{proof}
\end{lemma}

Using this value function, the optimal allocation thresholds can be obtained using the procedure provided in the following theorem.

\begin{figure}[t]
	\centering
	\resizebox {0.5\columnwidth} {!} {
		\begin{tikzpicture}
		[
		dot/.style={circle,draw=black, fill,inner sep=1pt}
		]
		
		\foreach \x in {1,...,4}
		\draw (\x,.0) -- node[below,yshift=-1mm] {\x} (\x,-.0);
		\node[below,yshift=-1mm] at (5,0) {5};
		\node[below,xshift=-2mm,yshift=-1mm] at (0,0) {0};
		\foreach \y in {1,...,5}
		\draw (.1,\y) -- node[below,xshift=-2mm] {\y} (-.1,\y);
		\node[below,xshift=-2mm] at (0,4) {4};
		\node[below,xshift=-2mm] at (0,5) {5};
		\draw[->,thick,-latex] (0,-1) -- (0,6);
		\draw[->,thick,-latex] (-1,0) -- (6,0);
		\node[left,xshift=0mm] at (0,6) {$N$};
		\node[below,yshift=-2mm] at (6,0) {$T$};
		\foreach \x in {0,...,5}{
			\node[dot,line width=1mm,color = red,fill=red] at (\x,0){ };
		}
		\foreach \x in {0,...,5}{
			\foreach \y in {\x}{
				\node[dot,line width=1mm,color = red,fill = red] at (\x,\y){ };
			}
		}
		\foreach \x in {2,...,5}{
			\node[dot,line width=1.3mm,color = blue] at (\x,1){ };
		}
		\foreach \x in {3,...,5}{
			\node[dot,line width=1.3mm,color = blue] at (\x,2){ };
		}
		\foreach \x in {4,...,5}{
			\node[dot,line width=1.3mm,color = blue] at (\x,3){ };
		}
		\node[dot,line width=1.3mm,color = blue] at (5,4){ };
		\end{tikzpicture}
	}
	\caption{Red dots indicate the boundary cases for the pair ($T,N$). Blue dots indicate the cases for which the decision problem needs to be solved for allocation.
		\vspace{-0.1in}}
	\label{Fig:grid}
\end{figure}
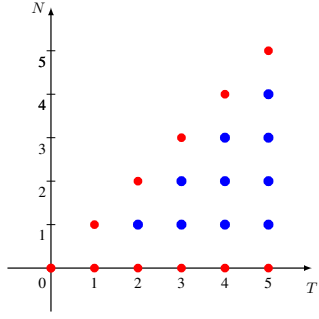

\begin{theorem} \label{Main_th}
If there are $T$ allocation periods and $N$ homogeneous resources available, then it is optimal to allocate an available resource to a utility maximizing request, characterized by $\tilde{Z}_T$, i.e., $q_T = 1$, if
\begin{align}
\tilde{Z}_T \geq \rho_T^n,
\end{align}
where the thresholds $\rho_T^n$ can be obtained as
\begin{align}
\rho_T^n = \mathbb{E}[V(T-1,n)] - \mathbb{E}[V(T-1,n-1)],
\end{align}
and the expected value functions with $T$ allocation periods and $N$ resources can be computed recursively as follows:
\begin{align} \label{E_V_computation}
\mathbb{E}[V(T,n)] &=   \int_{\rho_T^n}^{\infty}    \left(     \tilde{Z} + \mathbb{E}[V(T-1,n-1)]  \right)  f_{\tilde{Z}}(z) \ dz     +   \notag \\  &  \mathbb{E}[V(T-1,n)] F_{\tilde{Z}}(\rho_T^n).
\end{align}
\begin{proof}
	See \textbf{Appendix~\ref{Proof_main_th}}.
\end{proof}
\end{theorem}
To complete the optimal recursive solution, we need to evaluate the boundary conditions, i.e., the expected value functions for the ($T,N$) pairs highlighted by the red dots in Fig.~\ref{Fig:grid}. The expected values for such cases can be expressed by the following lemma.

\begin{lemma}
	The expected value obtained for the boundary cases of $(T,N)$ can be evaluated as follows.
	\begin{align}
	&\mathbb{E}[ V(T,0) ] = 0, \\
	&\mathbb{E}[ V(n,n) ] = n \mathbb{E}[\tilde{Z}].
	\end{align}
	\begin{proof}
		It follows from Theorem 1 that $\rho_n^n = \mathbb{E}[V(n-1,n)] - \mathbb{E}[V(n-1,n-1)] = 0, \forall n \in \mathbb{Z}^+$. It means that if the number of allocation periods equal the number of available resources, then the resource should be allocated to the utility maximizing request during that period regardless of the utility. Using this fact and the definition of $V(T,N)$, the result for $\mathbb{E}[V(n,n)]$ can be proved inductively.
	\end{proof}
\end{lemma}

\vspace{-0.1in}
\subsection{Computation \& Implementation}
\begin{figure}[b]
	\centering
	\includegraphics[width=2.4in]{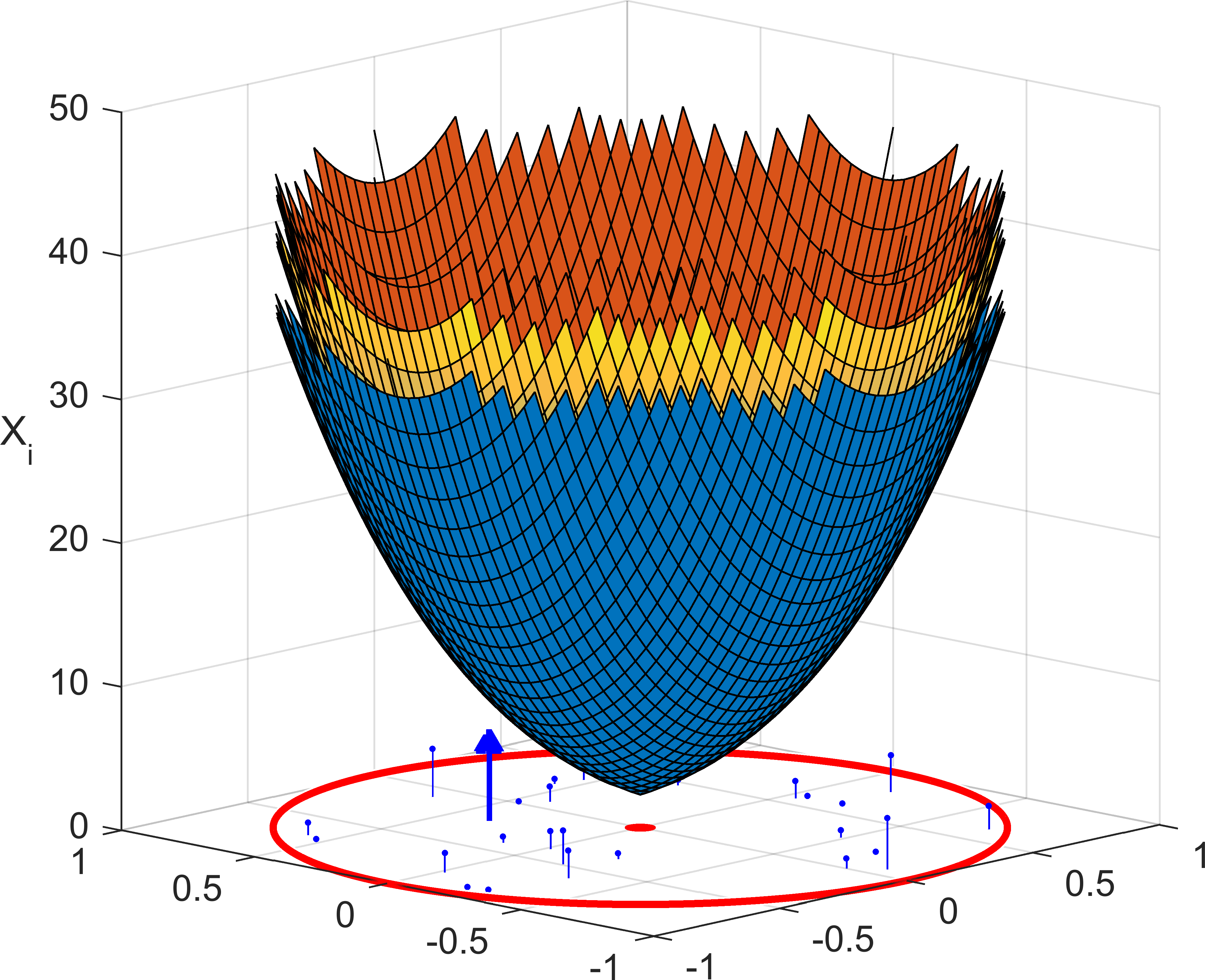}
	\caption{Spatial requests in one allocation period. The bold arrow represents the request with maximum intensity. Concentric surfaces correspond to the allocation thresholds if one, two, and three resources are available respectively.\vspace{-0.1in} \label{Thresh_Power_Exp_3d}}
\end{figure}

In this section, we explain the procedure to compute the optimal allocation thresholds and provide an overview on implementing the proposed framework.
In the case where there is one resource and two remaining time slots, the value function can be written as follows:
\begin{align}
V(T = 2, N = 1) =
\left\{
	\begin{array}{ll}
		\tilde{Z}  & \mbox{if  } \ \tilde{Z} \geq \rho_2^1, \\
		\rho_2^1 & \mbox{if  } \ \tilde{Z} < \rho_2^1.
	\end{array}
\right.
\end{align}
where the threshold $\rho_2^1 = \mathbb{E}[V(1,1)] = \mathbb{E}[\tilde{Z}]$.
In the case that there is one resource and $T = 3$ time slots remaining, then the value function can be expressed as follows:
\begin{align}
V(T = 3, N = 1) =
\left\{
	\begin{array}{ll}
		\tilde{Z}  & \mbox{if  } \ \tilde{Z} \geq \rho_3^1, \\
		\mathbb{E}[V(2,1)] & \mbox{if  } \ \tilde{Z} < \rho_3^1,
	\end{array}
\right.
\end{align}
where $\rho_3^1 = \mathbb{E}[V(2,1)] - \mathbb{E}[V(2,0)] = \int_{\rho_2^1}^{\infty} z f_{\tilde{Z}}(z) d z + \rho_2^1 F_{\tilde{Z}}(\rho_2^1)$. However, $\rho_2^1$ is available from the previous step. Similarly, by computing $\mathbb{E}[V(3,1)]$ enables the computation of $\rho_3^2$ and $V(3,2)$. This process needs to be executed recursively to obtain the value functions and the thresholds for arbitrary number of allocation periods and available resources. The step-wise procedure for computing the allocation thresholds is summarized in Algorithm~\ref{Alg:compute} and a summary of the implementation procedure is provided in Algorithm~\ref{Alg:implement}. Once, the form of $U(X,D)$ is known, the optimal thresholds $\rho_T^N$ translate to concentric surfaces, as shown in Fig.~\ref{Thresh_Power_Exp_3d}, that can be used to compare the maximal utilities in each time slot based on the number of available resources and time slots remaining to decide on allocation.

\begin{algorithm}[t]
  \small
  \caption{Optimal Threshold Computation}
  \begin{algorithmic}[1]
  \Procedure{Threshold Computation}{}
  \Require{$F_Z(z), f_Z(z), \mathbb{E}[\tilde{Z}]$.}
  \Initialize{$EV \gets [\boldsymbol{0}_{T \times N}], \rho \gets [\boldsymbol{0}_{T \times N}]$.}
  \For{$i = 1$ to $T$}
    \For{$j = 1$ to $N$}
      \If{$i=j$}
        \State $EV(i,j) \gets j \times \mathbb{E}[\tilde{Z}]$.
      \EndIf
    \EndFor
  \EndFor
  \For{$i = 1$ to $T$}
    \For{$j = 1$ to $N$}
    \If{$j = 1$}
        \State $\rho(i,j) \gets EV(i-1,j)$.
        \State $EV(i,j) = \int_{\rho(i,j)}^{\infty}  \tilde{Z} f_{\tilde{Z}}(z) \ dz  +$ \newline \qquad $EV(i-1,j) F_{\tilde{Z}}(\rho(i,j))$.
    \Else
        \State $\rho(i,j) \gets EV(i-1,j) - EV(i-1,j-1)$.
        \State Compute $EV(i,j)$ using~\eqref{E_V_computation}.
      \EndIf
    \EndFor
  \EndFor
  \EndProcedure
  \end{algorithmic}
  \label{Alg:compute}
  \end{algorithm}

\vspace{-0.1in}

\section{Results \& Discussion} \label{Sec:Results}
In this section, we first present some special cases for which the optimal solution can be obtained numerically. Then we provide simulation results to demonstrate the practical implementation and effectiveness of the proposed framework.
For sake of simple presentation of results, we assume a homogeneous spatio-temporal intensity of service requests, i.e., $\lambda(r,\theta,t) = \tau \lambda$. Then, service requests are distributed according to a Poisson process with intensity $\Lambda = \int_0^{2 \pi} \int_0^R \lambda \tau r  {\rm d}  r {\rm d} \theta = \tau \lambda \pi R^2$. For analytical tractability and practical relevance, we will use two specific form of utility functions, i.e., $U(X,D) = X (1+D)^{-\eta}$, $\eta \geq 0$, referred to as the \emph{power law utility}\footnote{Power law models are commonly used to model the propagation of wireless signals.} and $U(X,D) = X e^{- \alpha D}$, referred to as the \emph{exponential utility}.

\subsection{Special Cases}

\subsubsection{Power Law Utility}
In this section, we assume that the utility function is of the form of $U(X,D) = X (1+D)^{-\eta}$, $\eta \geq 0$. We will further break down into two special cases, i.e., when the intensities of requests are exponentially and uniformly distributed.

 \textbf{Case I: Exponential Intensity}\\
If $X \sim \text{ Exp}(\mu^{-1})$, then $F_X(x) = 1 - e^{-\mu x}, x \geq 0$, and $f_X(x) = \mu e^{-\mu x}, x \geq 0$, then the pdf and cdf of the utility of each service request can be expressed by the following corollary.

\begin{corollary} \label{Power_Exp}
The pdf and cdf of the utility $Z$ of a service request inside a circular region of radius $R$ if $U(X,D) = X(1+D)^{-\eta}$ and $X \sim {\rm Exp}(\mu^{-1})$, can be evaluated as follows:
\begin{align}
F_Z(z) &= 1 - \frac{2}{\eta R^2} \left( E_{\frac{\eta - 2}{\eta}}(z \mu)  - \right. \notag \\
& \left. (1+R)^2 E_{\frac{\eta - 2}{\eta}}(z \mu (1+R)^{\eta}) - \right. \notag \\
& \left. \quad E_{\frac{\eta - 1}{\eta}}(z \mu) + (1+R) E_{\frac{\eta - 1}{\eta}}(z \mu (1+R)^{\eta}) \right),
\end{align}
where $E_n(x) = \int_{1}^{\infty} \frac{e^{-xt}}{t^{\eta}} dt$ is the generalized exponential integral.
\begin{align}
f_Z(z) = \int_{0}^{R} \mu d (1+d)^{\eta} e^{-\mu z (1 + d)^{\eta}}    {\rm d} d.
\end{align}
\begin{proof}
See \textbf{Appendix~\ref{Proof_Power_Exp}}.
\end{proof}
\end{corollary}

\begin{algorithm}[t]
	\small
	\caption{Spatio-temporal Resource Allocation}
	\begin{algorithmic}[1]
		\Procedure{Runtime}{}
		\Require{$T,N,\rho$.}
		\While{$T \ge 0$}
		\State Obtain the tuple $(X_i,D_i)_{1\leq i \leq K}$.
		\State Compute the utility $U(X_i,D_i)$ for all requests.
		\State Determine the maximal utility $\tilde{Z}$ using~\eqref{maximal_utility}.
		\If{$\tilde{Z}\ge \rho(T,N)$}
		\State Allocate a resource to request corresponding to $\tilde{Z}$.
		\State $N \gets N-1$
		\Else
		\State Skip allocation in the current allocation period.
		\EndIf
		\State $T \gets T-1$
		\EndWhile
		\EndProcedure
	\end{algorithmic}
	\label{Alg:implement}
\end{algorithm}

%
%


 \textbf{Case II: Uniform Intensity}\\
If $X \sim \text{ Unif}(0,\beta)$, then $F_X(x) = \frac{x }{\beta}, \ \ x \in (0,\beta)$ and $f_X(x) = \frac{1}{\beta}, \ \ x \in (0,\beta)$, then the pdf and cdf of the utility of each service request can be expressed by the following corollary.

\begin{corollary} \label{Power_Unif}
The pdf and cdf of the utility $Z$ of a service request inside a circular region of radius $R$ if $U(X,D) = X(1+D)^{-\eta}$ and $X \sim {\rm U}(0, \beta)$, can be evaluated as follows:
\begin{align}
F_Z(z) = \left\{
\begin{array}{ll}
\frac{2z}{\beta R^2} \left( \frac{1 + \left( \left( \frac{\beta}{z}\right)^{1/\eta} \right)^{\eta + 1}(R(\eta + 1) - 1)}{2 + 3 \eta + \eta^2}  \right),  \notag \\
\qquad  \qquad \qquad  \qquad \qquad  \mbox{if } 0 \leq z \leq \beta(1+R)^{-\eta} \\
\frac{2z}{\beta R^2} \left( \frac{1 + (R+1)^{\eta + 1}((\eta + 1)\left(\left( \frac{\beta}{z}\right)^{(1/\eta)}\right) - \eta - 2)}{2 + 3 \eta + \eta^2}  \right) + \notag \\
\frac{2}{R^2}\left(  \frac{R^2}{2} - \frac{1}{2} \left( \left(\frac{\beta}{z}\right)^{1/\eta} - 1\right)^2  \right), \\  \qquad  \qquad \qquad  \qquad  \mbox{if } \beta(1+R)^{-\eta} < z \leq \beta
\end{array}
\right.
\end{align}
and
\begin{align}
f_Z(z) = \left\{
\begin{array}{ll}
\frac{2}{\beta R^2} \left( \frac{1 + (R+1)^{\eta + 1}(R(\eta + 1) - 1)}{2 + 3 \eta + \eta^2}  \right),  \\
\qquad  \qquad \qquad  \qquad \qquad \mbox{if } 0 \leq z \leq \beta(1+R)^{-\eta} \\
\frac{1}{\beta R^2 \eta} \left(  2\eta \left( 1  -  2 \left(  \frac{\beta}{z} \right)^{1 + 2/\eta}  +  \left( \frac{\beta}{z} \right)^{\frac{\eta + 1}{\eta}} \right)  \right. - \\
\left. 4 \left( \frac{\beta}{z} \right)^{\frac{\eta + 1}{\eta}}   \left(  \left( \frac{\beta}{z} \right)^{1/\eta} - 1 \right) \right), \\
\qquad  \qquad \qquad  \quad \mbox{if } \beta(1+R)^{-\eta} < z \leq \beta
\end{array}
\right.
\end{align}
\begin{proof}
See \textbf{Appendix~\ref{Proof_Power_Unif}}.
\end{proof}
\end{corollary}


%

\subsubsection{Exponential Utility}
In this section, we assume that the utility function is of the form of $U(X,D) = X e^{-\alpha D}$, $\alpha \geq 0$. We will further break down into two special cases, i.e., when the intensities of requests are exponentially and uniformly distributed.

 \textbf{Case I: Exponential Intensity}\\
If the intensity of requests follows an exponential distribution with mean $\mu^{-1}$, then the pdf and cdf of the utility of each service request can be expressed by the following corollary.
\begin{corollary} \label{Exp_Exp}
The pdf and cdf of the utility $Z$ of a service request inside a circular region of radius $R$ if $X e^{-\alpha D}$, $\alpha \geq 0$ and $X \sim {\rm Exp}(\mu^{-1})$, can be evaluated as follows:
\begin{align}
F_Z(z) &=1 - \frac{2}{R^2} \int_{0}^{R} d e^{-\mu z e^{\alpha d}} {\rm d} d.
\end{align}
and
\begin{align}
f_Z(z) = \int_{0}^{R} \frac{2\mu d}{R^2}  e^{\alpha d -\mu z e^{\alpha d}}   {\rm d} d.
\end{align}
\begin{proof}
The proof follows a similar methodology as used in~\textbf{Appendix~\ref{Proof_Power_Exp}} and has been omitted for brevity.
\end{proof}
\end{corollary}
%

 \textbf{Case II: Uniform Intensity}\\
If the intensity of requests follows a uniform distribution in $(0,\beta)$, then the pdf and cdf of the utility of each service request can be expressed by the following corollary.
\begin{corollary} \label{Exp_Unif}
The pdf and cdf of the utility $Z$ of a service request inside a circular region of radius $R$ if $X e^{-\alpha D}$, $\alpha \geq 0$ and $X \sim {\rm U}(0,\beta)$, can be evaluated as follows:
\begin{align}
F_Z(z) &= \left\{
\begin{array}{ll}
\frac{2z}{\alpha^2 \beta R^2 } \left( 1 + e^{\alpha R} (\alpha R - 1)  \right),  \\
  \qquad  \qquad \qquad  \qquad \qquad   \qquad \mbox{if } 0 \leq z \leq \beta e^{-\alpha R} \\
\frac{2}{\alpha^2 \beta R^2 } \left(   z  +  ze^{\alpha R} (\alpha R - 1)  \right)
, \\  \qquad  \qquad \qquad  \qquad \qquad \qquad \mbox{if } \beta e^{-\alpha R} < z \leq \beta
\end{array}
\right.
\end{align}
\begin{align}
f_Z(z) = \left\{
\begin{array}{ll}
\frac{2}{\alpha^2 \beta R^2 } \left( 1 + e^{\alpha R} (\alpha R - 1)  \right),  \\
\qquad \qquad \qquad \qquad \qquad \quad \mbox{if } 0 \leq z \leq \beta e^{-\alpha R} \\
\frac{2}{\alpha^2 \beta R^2 } \left(  1      + e^{\alpha R}(\alpha R - 1) \right) , \\  \qquad  \qquad \qquad \quad \qquad \qquad  \mbox{if } \beta e^{-\alpha R} < z \leq \beta
\end{array}
\right.
\end{align}
\begin{proof}
The proof follows a similar methodology as used in~\textbf{Appendix~\ref{Proof_Power_Unif}} and has been omitted for brevity.  
\end{proof}
\end{corollary}

\begin{figure}[t]
\centering
\subfloat[Power law utility with exponential intensity.]{\includegraphics[width=1.6in]{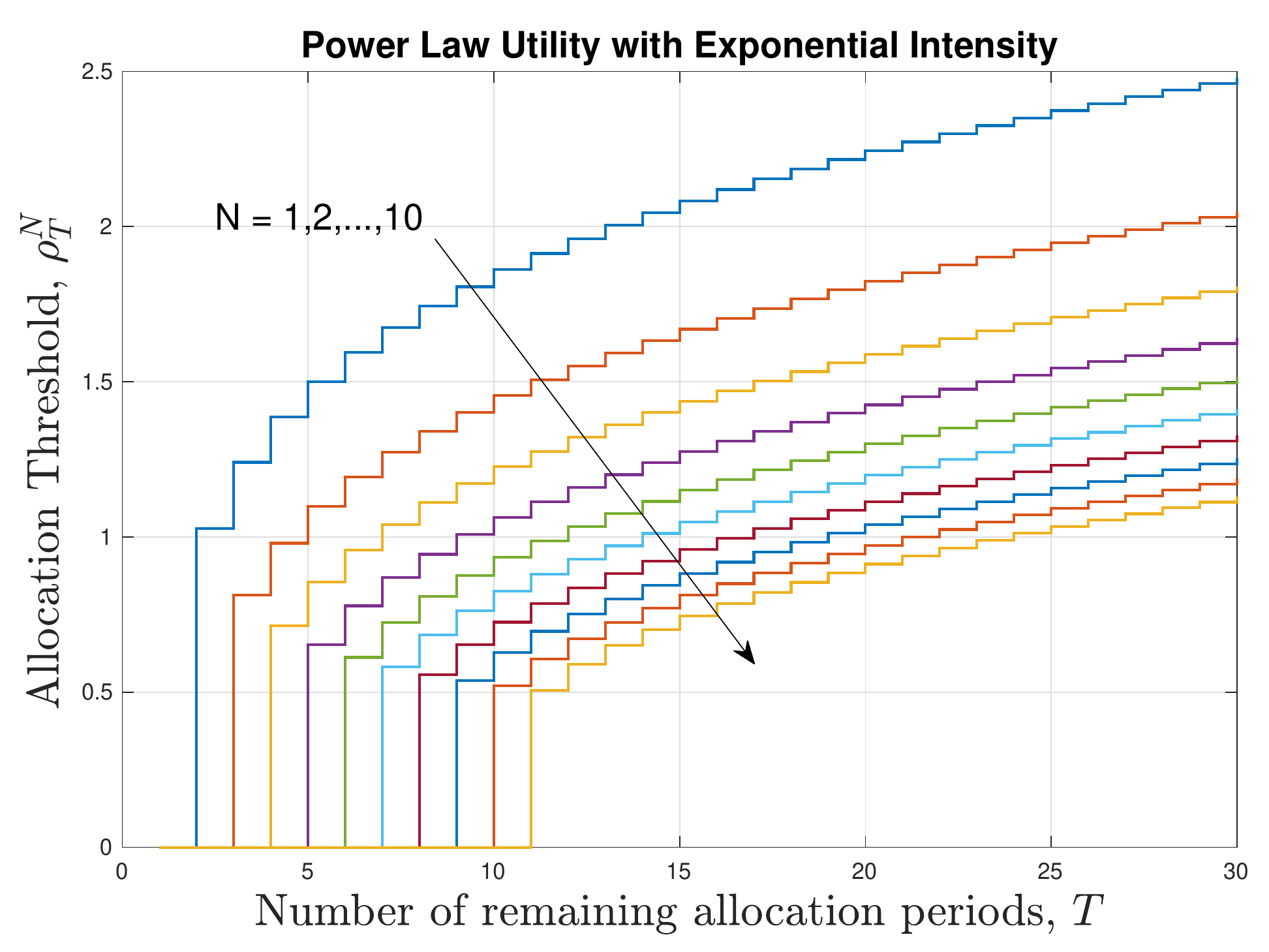} \label{Thresh_Power_Exp}} \ \
\subfloat[Exponential utility with exponential intensity.]{\includegraphics[width=1.6in]{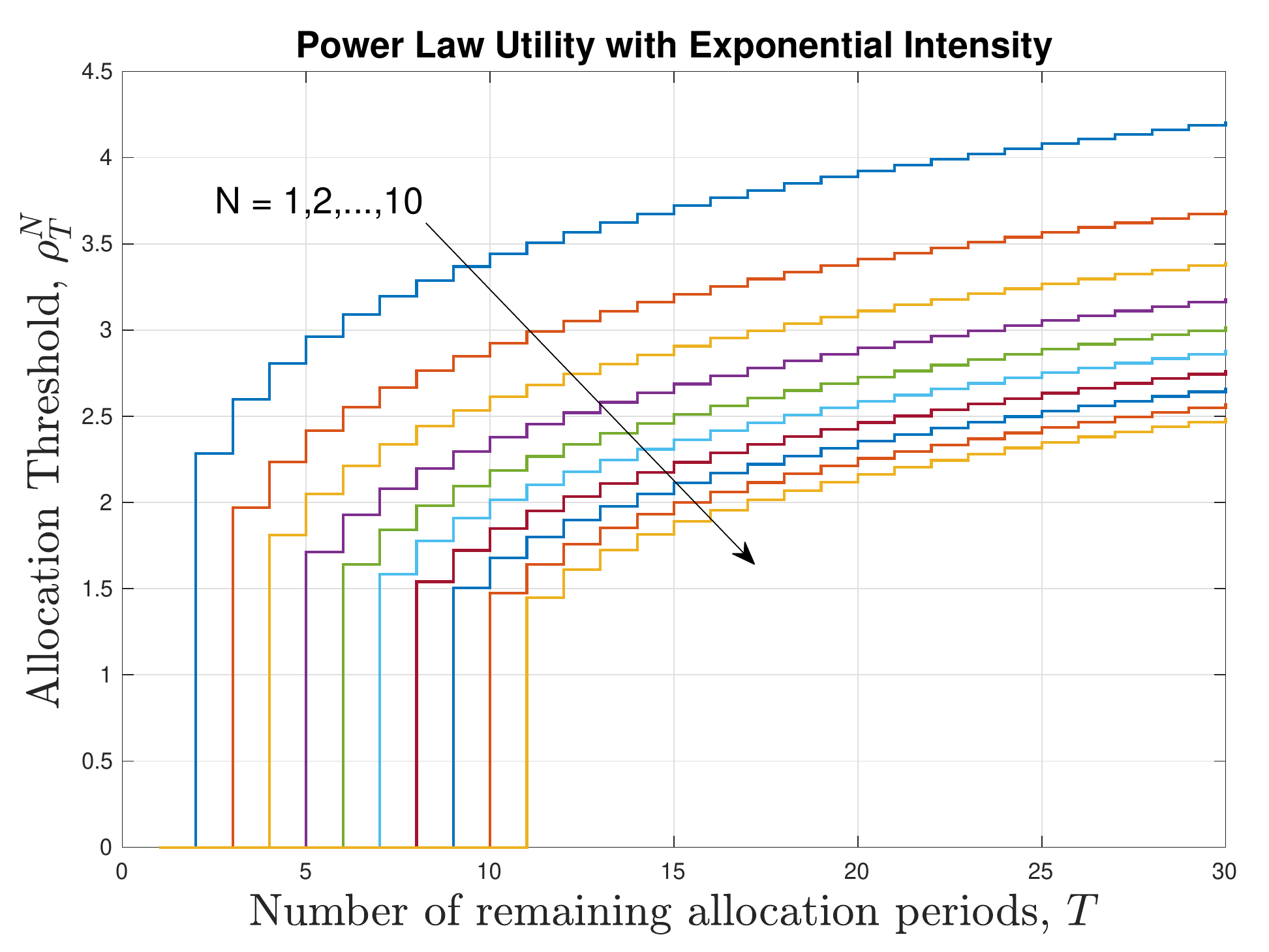} \label{Thresh_Exp_Exp}}
\caption{Resource allocation thresholds for exponential intensity of service requests.\vspace{-0.1in}}
\label{Fig:Thresh_X_Exp}
\end{figure}


\subsection{Performance Evaluation \& Comparison}
To illustrate the performance of our developed resource provisioning framework, we conduct simulation experiments. The model parameters, unless otherwise stated, are selected as follows for illustrative purposes: service range $R = 1$, service request density $\lambda = 10$ requests per unit area, duration of each time slot $\tau$ = 1, number of time slots $T = 30$, number of available resources $N = 10$, mean of exponential intensity $\mu^{-1} = 1$, parameter of uniform intensity $ \beta = 1$. Service requests are generated according to the homogeneous spatio-temporal Poisson process in a circular region around the origin. At each successive time slot, the maximum utility is picked and a resource is allocated to the corresponding request only if the utility is higher than the threshold. Otherwise the requests are discarded and the next time slot is observed until the terminal time is reached.

Fig.~\ref{Fig:Thresh_X_Exp} plots the optimal allocation thresholds against the number of allocation periods remaining for different number of available resources at the source node if the intensity is assumed to be exponentially distributed. The threshold successively increase as more allocation periods are available for the same number of available resources. In other words, the source becomes more selective in allocation as more allocation periods are available. Similarly, for a given number of allocation periods, the threshold decreases as the the number of available resources increase. Note that the threshold remains zero if the number of available resources and remaining allocation periods are equal. The successive difference between the thresholds is attributed to the fact that the intensity of requests is exponentially distributed.
A similar behavior will be observed in the case of uniformly distributed arrivals except that the difference between successive thresholds is expected to be uniform.

\begin{figure}
	\centering
	\includegraphics[width=3in]{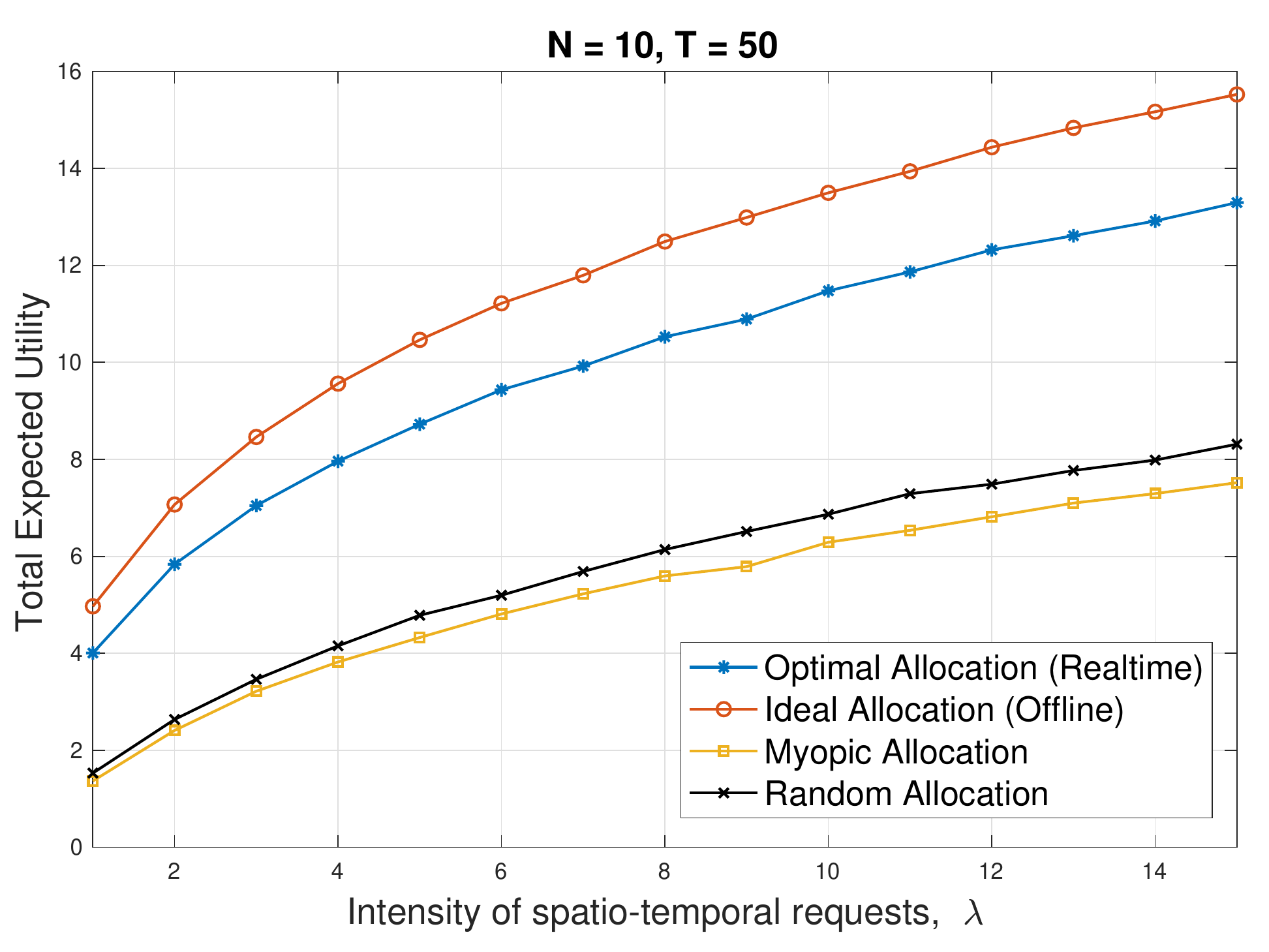}
	\caption{Total expected utility against varying spatio-temporal density of requests.}
\label{Fig:comparison}
\end{figure}
\begin{figure}
	\centering
	\includegraphics[width=3in]{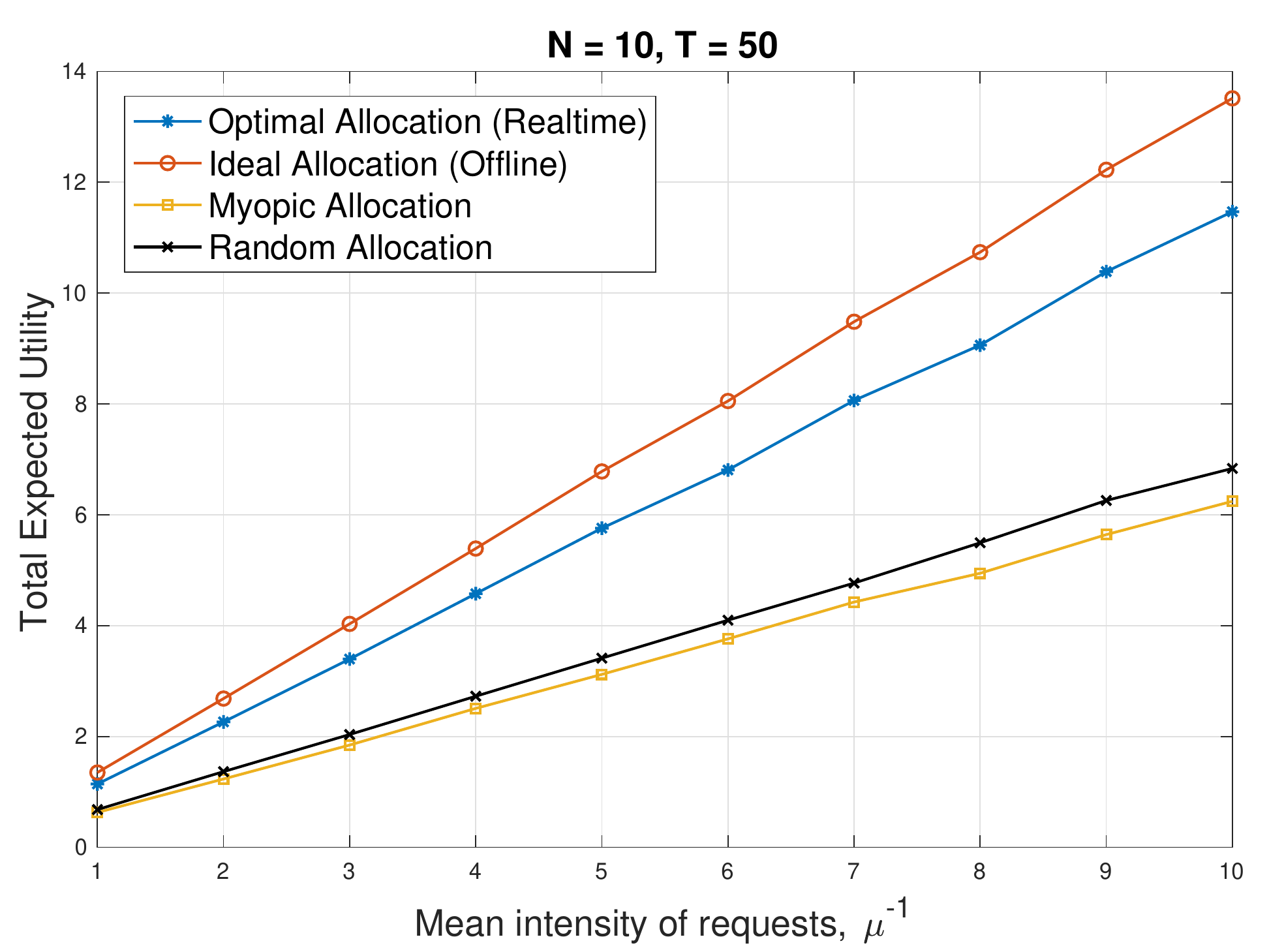}
	\caption{Total expected utility against varying intensity of requests. \vspace{-0.1in}}
\label{Fig:comparison_mu}
\end{figure}

We also compare our proposed resource provisioning framework with benchmark schemes namely the Ideal allocation, myopic allocation, and random allocation. The Ideal allocation represents the case when there is no uncertainty about the future and the maximum utilities in each time slot are known a priori. Therefore, it results in the best possible resource allocation strategy. The Myopic allocation strategy allocates a resource to the utility maximizing request in each time slot regardless of its magnitude. The random allocation is a generalization of the myopic allocation where in every time slot, a resource is allocated to the utility maximizing request with a probability of 0.5. If the probability of allocation is further reduced, there may be an increase in total expected utility since there is a higher chance of hitting better utility requests.
Fig.~\ref{Fig:comparison} and Fig.~\ref{Fig:comparison_mu} plot the total expected utility of allocation under varying intensity of spatio-temporal requests and the mean intensity of requests respectively. The simulation results clearly show that the proposed optimal allocation strategy lower expected utility as compared to the Ideal allocation. The loss in utility is due to the fact that allocation decisions are made in real-time without information about future arrival of requests. However, it performs at least two-fold better than the myopic and random allocation strategies.

\section{Conclusions \& Future Work} \label{Sec:Conclusion}
In this paper, we have proposed a utility maximizing approach to allocate resources from a centralized source location to spatio-temporal service requests with varying intensity. The framework is highly generic in terms of the utility of allocation, the number of available resources and the number of allocation periods along with the spatio-temporal statistics of the requests. Using statistical analysis of the utility obtained by the allocation, we have developed an optimal filtering scheme that makes only qualifying requests eligible for resource allocation. The developed resource provisioning framework is envisioned to have wide ranging applications in smart cities.


Several useful extensions can be done to enhance the proposed framework to cater for more realistic and tailored scenarios for different resource allocation problems. The cost of delaying the allocation can be added as a parameter to the framework to prevent over-selectiveness of the source. Furthermore, the case of non-homogeneous resources and the continuous time version of the framework can be investigated
to cater for more versatile scenarios and applications.

\appendices

\section{Proof of Lemma~\ref{dist_pdf}}\label{dist_pdf_proof}
The cdf of the distance to a typical service request in a non-homogeneous PPP model with density $\tau \tilde{\lambda}(r,\theta)$ can be evaluated as follows:
\begin{align}
F_{D}(d) = \mathbb{P}(D \leq d) = \frac{  \int_0^{2 \pi} \int_0^d \tilde{\lambda}(r,\theta) r {\rm d} r  {\rm d} \theta }{ \int_0^{2 \pi} \int_0^R \tilde{\lambda}(r,\theta) r  d r  d \theta}. \notag
\end{align}
The pdf can then be obtained as follows:
\begin{align}
f_{D}(d) = \frac{ {\rm d} F_{D}(d)}{ {\rm d} d} = \frac{\int_0^{2 \pi} d \tilde{\lambda}(d,\theta)   {\rm d} \theta}{\int_0^{2 \pi} \int_0^R \tilde{\lambda}(r,\theta) r  {\rm d} r  {\rm d} \theta}. \notag
\end{align}
In the special case of a homogeneous PPP, i.e., $\tilde{\lambda}(r,\theta) = \tau \lambda$, the cdf and pdf of the distance can be expressed as follows:
\begin{align}
F_{D}(d) &= \mathbb{P}(D \leq d) = \frac{\tau \lambda \pi d^2}{ \tau \lambda \pi R^2} = \frac{d^2}{R^2},
\end{align}
\begin{align}
f_{D}(d) &= \frac{ d F_{D}(d)}{ {\rm d} d} = \frac{2d}{R^2}.
\end{align}

\section{Proof of Lemma~\ref{max_pdf_lemma}}\label{max_pdf_lemma_proof}
The distribution function of $\tilde{Z} = \max \{ Z_1, Z_2, \ldots, Z_K\}$ with $K$ being a Poisson random variable can be obtained as follows:
\begin{align}
F_{\tilde{Z}}(z|K=k) &= \mathbb{P}(\tilde{Z} \leq z | K = k), \notag \\
&= \mathbb{P}(\max \{Z_1, Z_2, \ldots, Z_K \} \leq z | K = k ), \notag \\
&= \mathbb{P}(Z_1 \leq z, Z_2 \leq z, \ldots, Z_K \leq z | K = k), \notag\\
&= \prod_{i=1}^{K} \mathbb{P}(Z_i \leq z) = \prod_{i=1}^{K} F_{\tilde{Z}}(z)= (F_Z(z))^K.
\end{align}
Consequently, the conditional pdf of $\tilde{Z}$ can be expressed as follows:
\begin{align}
f_{\tilde{Z}}(z|K = k) = \frac{dF_{\tilde{Z}}(z|K = k)}{dz} = k(F_Z(z))^{k-1}f_Z(z).
\end{align}
Finally, the pdf can be obtained as follows:
\begin{align}
f_{\tilde{Z}}(z) &= \sum_{k = 1}^{\infty} f_{\tilde{Z}}(z|K = k) \mathbb{P}(K = k | K > 0), \notag \\
&= \sum_{k = 1}^{\infty} k(F_Z(z))^{k-1}f_Z(z) \frac{e^{- \Lambda } (\Lambda )^k }{(1 - e^{-\Lambda })k!}, \notag \\
&= \frac{\Lambda  f_Z(z) e^{\Lambda  (F_Z(z) - 1)}}{1 - e^{-\Lambda }}.
\end{align}

\section{Proof of Lemma~\ref{Value_lemma}}\label{Proof_value_lemma}
\begin{figure}[h!]
	\centering
	\begin{adjustbox}{width=0.5\textwidth}
		\begin{tikzpicture}[level distance=1.5cm,
		level 1/.style={sibling distance=7cm},
		level 2/.style={sibling distance=4cm},scale = 0.7]
		\node {\begin{tabular}{c} $V(T,N)$ \end{tabular}}
		child {node {$\tilde{Z}$}
			child {node[solid] {$\mathbb{E}[V(T-1,N-1)]$} }
		}
		child {node {$0$}
			child {node[solid] {$\mathbb{E}[V(T-1,N)]$} }
		};
		\end{tikzpicture}
	\end{adjustbox}
	\caption{Decision tree at when $T$ allocation periods are remaining and $N$ resources are available.}
	\label{fig:decision_tree}
\end{figure}
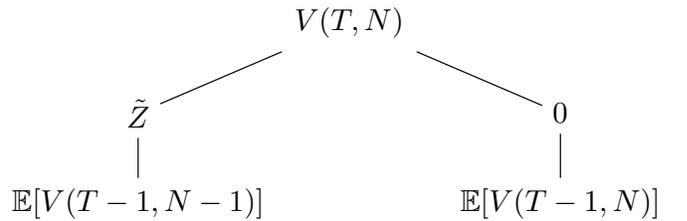

To prove the recursive value function, we make use of the decision tree shown in Fig.~\ref{fig:decision_tree}. At the current allocation period, if there are $T$ allocation periods remaining with $N$ available resources, there are two possible decisions, i.e., to allocate a resource to the utility maximizing request or to postpone allocation to a future allocation period.
The value function can be expressed as follows:
\begin{align}
&V(T,N) = \\ \notag & \underset{q_T}{\max} \{ \tilde{Z} q_T + \mathbb{E}[V(T-1,N-1) \ , \ \mathbb{E}[V(T-1,N)]\}.
\end{align}
It is clear that if $q_T=1$ and $\tilde{Z}_T$ is greater than some threshold $\rho_T^N$, then $\tilde{Z} + \mathbb{E}[V(T-1,N-1)] \geq \mathbb{E}[V(T-1,N)]$. Therefore, $V(T,N) = \tilde{Z} + \mathbb{E}[V(T-1,N-1)]$. Otherwise, if $q_T = 0$, then $\mathbb{E}[V(T-1,N)] > \mathbb{E}[V(T-1,N-1)]$, which results in $V(T,N) = \mathbb{E}[V(T-1,N)]$.


\vspace{-0.1in}
\section{Proof of Theorem~\ref{Main_th}}\label{Proof_main_th}
From the decision tree shown in Fig.~\ref{fig:decision_tree}, the decision to allocate a resource to a qualifying request is only made if the utility obtained from the allocation is higher than the utility obtained from postponing the decision. It implies that
\begin{align}
\tilde{Z}  + \mathbb{E}[V(T-1,n-1)] \geq \mathbb{E}[V(T-1,n)].
\end{align}
This leads to the condition that
\begin{align}
\tilde{Z} \geq \mathbb{E}[V(T-1,n)] - \mathbb{E}[V(T-1,n-1)] = \rho_T^N.
\end{align}
The expectation of the value function can be obtained directly from the definition in~\eqref{Value_lemma} in terms of the allocation threshold $\rho_T^N$.

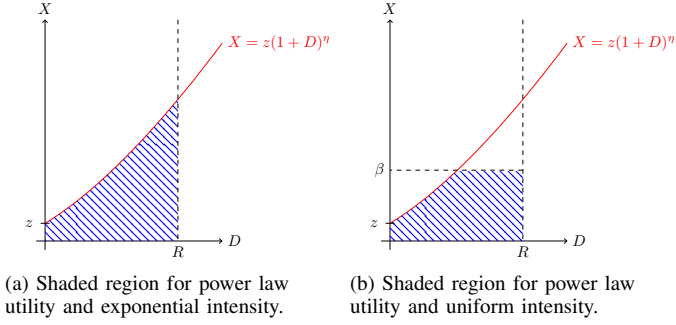
\begin{figure}[t]
\centering
\subfloat[Shaded region for power law\newline utility and exponential intensity.]{
\resizebox {0.5\columnwidth} {!} {
\begin{tikzpicture}[domain=0:4]
        \draw[->] (-0.2,0) -- (4,0) node[right] {$D$};
         \draw[->] (0,-0.2) -- (0,5) node[above] {$X$};
        \draw[color=red] plot (\x, { 0.4*(1+\x)^1.5 })
        node[right] {$X = z(1 + D)^{\eta}$};
        \path[pattern= north west lines, pattern color = blue] plot[smooth, samples=100, domain=0:3] (\x,{ 0.4*(1+\x)^1.5 }) -| (3,0) -| (0,0) -- cycle;
        \draw[dashed] (3,5) -- (3,0) node [below]{$R$};
        \draw[solid] (-0.1,0.4) -- (0.1,0.4) node [left]{$z \ \ $};
        \end{tikzpicture}
        }
        \label{power_law_region}
}
\subfloat[Shaded region for power law\newline utility and uniform intensity.]{
	\resizebox {0.5\columnwidth} {!} {
		\begin{tikzpicture}[domain=0:4]
		\draw[->] (-0.2,0) -- (4,0) node[right] {$D$};
		\draw[->] (0,-0.2) -- (0,5) node[above] {$X$};
		\draw[color=red] plot (\x, { 0.4*(1+\x)^1.5 })
		node[right] {$X = z(1 + D)^{\eta}$};
		\path[pattern= north west lines, pattern color = blue] plot[smooth, samples=100, domain=0:1.5] (\x,{ 0.4*(1+\x)^1.5 }) -| (3,0) -| (0,0) -- cycle;
		\draw[dashed] (3,5) -- (3,0) node [below]{$R$};
		\draw[dashed] (3,1.6) -- (0,1.6) node [left]{$\beta$};
		\draw[solid] (-0.1,0.4) -- (0.1,0.4) node [left]{$z \ \ $};
		\end{tikzpicture}
	}
	\label{power_law_region_unif}
}
\caption{The shaded region represents the region of integration $\mathcal{S} = \{(x,d): U(x,d) \leq z\}$. \vspace{-0.1in}}
\label{regions}
\end{figure}

\vspace{-0.1in}
\section{Proof of Corollary~\ref{Power_Exp}}\label{Proof_Power_Exp}
The pdf of $Z$ can be evaluated as follows:
\begin{align}
F_Z(z) &= \mathbb{P}[Z \leq z] = \mathbb{P}[X (1+D)^{-\eta} \leq z], \notag \\
&= \mathbb{P}[X \leq z(1+D)^{\eta}].
\end{align}
This probability can be obtained by integrating the joint density $f_{X,D}(x,d)$ over the shaded region $\mathcal{S} = \{(x,d): U(x,d) \leq z\}$ shown in Fig.~\ref{power_law_region}.
Consequently, the cdf can be expressed as follows:
\begin{align}
F_Z(z) &= \int_{0}^{R}  \int_{0}^{z(1+d)^\eta} f_X(x) f_{D}(d) {\rm d} x {\rm d} d, \notag \\
&= \int_{0}^{R}  \int_{0}^{z(1+d)^\eta} \mu e^{-\mu x} \times 2 \frac{d}{R^2} {\rm d} x {\rm d} d.
\end{align}
Computing the integrals results in the expression provided in Corollary~\ref{Power_Exp}. Furthermore, differentiating $F_Z(z)$ w.r.t. $z$ results in $f_Z(z)$.

\vspace{-0.2in}
\section{Proof of Corollary~\ref{Power_Unif}}\label{Proof_Power_Unif}
The pdf of $Z$ can be evaluated as follows:
\begin{align}
F_Z(z) &= \mathbb{P}[Z \leq z] = \mathbb{P}[X (1+D)^{-\eta} \leq z], \notag \\
&= \mathbb{P}[X \leq z(1+D)^{\eta}].
\end{align}
This probability can be obtained by integrating the joint density $f_{X,D}(x,d)$ over the shaded region $\mathcal{S} = \{(x,d): U(x,d) \leq z\}$ shown in Fig.~\ref{power_law_region_unif}.
\begin{align}
F_Z(z) &= \left\{
\begin{array}{ll}
 \int_{0}^{R}  \int_{0}^{z(1+d)^\eta} \frac{1}{\beta} \times 2 \frac{d}{R^2} {\rm d} x {\rm d}d,  \  \mbox{if } z \leq \beta(1+R)^{-\eta} \\
\int_{0}^{\left(\frac{\beta}{z}\right)^{\frac{1}{\eta}} - 1}  \int_{0}^{z(1+d)^\eta} \frac{1}{\beta} \times 2 \frac{d}{R^2} {\rm d} x {\rm d} d + \\ \int_{\left(\frac{\beta}{z}\right)^{\frac{1}{\eta}} - 1}^{R}  \int_{0}^{\beta} \frac{1}{\beta} \times 2 \frac{d}{R^2} {\rm d} x {\rm d} d, \\  \qquad  \qquad  \quad \mbox{if } \beta(1+R)^{-\eta} < z \leq \beta
\end{array}
\right.
\end{align}

Computing the integrals leads to the result in Lemma~\ref{Power_Unif}.
The density function of $Z$ can then be evaluated by differentiating $F_Z(z)$ w.r.t. $z$.


\vspace{-0.0in}
\bibliographystyle{IEEEtran}
\bibliography{references}

\begin{thebibliography}{10}
\providecommand{\url}[1]{#1}
\csname url@samestyle\endcsname
\providecommand{\newblock}{\relax}
\providecommand{\bibinfo}[2]{#2}
\providecommand{\BIBentrySTDinterwordspacing}{\spaceskip=0pt\relax}
\providecommand{\BIBentryALTinterwordstretchfactor}{4}
\providecommand{\BIBentryALTinterwordspacing}{\spaceskip=\fontdimen2\font plus
\BIBentryALTinterwordstretchfactor\fontdimen3\font minus
  \fontdimen4\font\relax}
\providecommand{\BIBforeignlanguage}[2]{{%
\expandafter\ifx\csname l@#1\endcsname\relax
\typeout{** WARNING: IEEEtran.bst: No hyphenation pattern has been}%
\typeout{** loaded for the language `#1'. Using the pattern for}%
\typeout{** the default language instead.}%
\else
\language=\csname l@#1\endcsname
\fi
#2}}
\providecommand{\BIBdecl}{\relax}
\BIBdecl

\bibitem{socially_responsible}
H.-R.~D. Tsai, Y.~Shoukry, M.~K. Lee, and V.~Raman, ``Towards a socially
  responsible smart city: Dynamic resource allocation for smarter community
  service,'' in \emph{Proceedings of the 4th ACM International Conference on
  Systems for Energy-Efficient Built Environments}, ser. BuildSys '17.\hskip
  1em plus 0.5em minus 0.4em\relax New York, NY, USA: ACM, 2017, pp.
  13:1--13:4.

\bibitem{city_of_things}
J.~Santos, T.~Vanhove, M.~Sebrechts, T.~Dupont, W.~Kerckhove, B.~Braem, G.~V.
  Seghbroeck, T.~Wauters, P.~Leroux, S.~Latre, B.~Volckaert, and F.~D. Turck,
  ``City of things: Enabling resource provisioning in smart cities,''
  \emph{IEEE Communications Magazine}, vol.~56, no.~7, pp. 177--183, July 2018.

\bibitem{resource_provisioning}
J.~Santos, T.~Wauters, B.~Volckaert, and F.~D. Turck, ``Resource provisioning
  for {IoT} application services in smart cities,'' in \emph{13th International
  Conference on Network and Service Management (CNSM 2017)}, Nov 2017, pp.
  1--9.

\bibitem{Villani_OT}
C.~Villani, \emph{{Optimal Transport: Old and New}}, ser. Grundlehren der
  mathematischen Wissenschaften.\hskip 1em plus 0.5em minus 0.4em\relax
  Springer, Sep 2008.

\bibitem{hungarian}
H.~W. Kuhn and B.~Yaw, ``The hungarian method for the assignment problem,''
  \emph{Naval Res. Logist. Quart}, pp. 83--97, 1955.

\bibitem{junaid_hungarian}
H.~Ghazzai, M.~J. Farooq, A.~Alsharoa, E.~Yaacoub, A.~Kadri, and M.~Alouini,
  ``Green networking in cellular hetnets: A unified radio resource management
  framework with base station {ON/OFF} switching,'' \emph{IEEE Transactions on
  Vehicular Technology}, vol.~66, no.~7, pp. 5879--5893, Jul. 2017.

\bibitem{taxi_dispatch}
F.~Miao, S.~Han, S.~Lin, Q.~Wang, J.~A. Stankovic, A.~Hendawi, D.~Zhang, T.~He,
  and G.~J. Pappas, ``Data-driven robust taxi dispatch under demand
  uncertainties,'' \emph{IEEE Transactions on Control Systems Technology},
  vol.~27, no.~1, pp. 175--191, Jan. 2019.

\bibitem{data_collection_magazine}
I.~Ali, A.~Gani, I.~Ahmedy, I.~Yaqoob, S.~Khan, and M.~H. Anisi, ``Data
  collection in smart communities using sensor cloud: Recent advances,
  taxonomy, and future research directions,'' \emph{IEEE Communications
  Magazine}, vol.~56, no.~7, pp. 192--197, July 2018.

\bibitem{sequential_stochastic}
C.~Derman, G.~J. Lieberman, and S.~M. Ross, ``A sequential stochastic
  assignment problem,'' \emph{Management Science}, vol.~18, no.~7, pp.
  349--355, 1972.

\bibitem{data_collection}
Q.~Zhu, M.~Y.~S. Uddin, Z.~Qin, and N.~Venkatasubramanian, ``Upload planning
  for mobile data collection in smart community internet-of-things
  deployments,'' in \emph{IEEE International Conference on Smart Computing
  (SMARTCOMP 2016)}, May 2016, pp. 1--8.

\bibitem{network_resource_management}
D.~Sahinel, C.~Akpolat, F.~Sivrikaya, and S.~Albayrak, ``An agent-based network
  resource management concept for smart city services,'' in \emph{14th Annual
  Conference on Wireless On-demand Network Systems and Services (WONS 2018)},
  Feb 2018, pp. 129--132.

\bibitem{urban_sensing}
Q.~Zhu, M.~Y.~S. Uddin, N.~Venkatasubramanian, and C.~Hsu, ``Spatiotemporal
  scheduling for crowd augmented urban sensing,'' in \emph{IEEE Conference on
  Computer Communications (INFOCOM 2018)}, April 2018.

\bibitem{data_collection_queueing}
Y.~Han, Y.~Zhu, and J.~Yu, ``Utility-maximizing data collection in crowd
  sensing: An optimal scheduling approach,'' in \emph{12th Annual IEEE
  International Conference on Sensing, Communication, and Networking (SECON
  2015)}, June 2015, pp. 345--353.

\bibitem{junaid_acc}
M.~J. Farooq and Q.~Zhu, ``Adaptive and resilient revenue maximizing dynamic
  resource allocation and pricing for cloud-enabled {IoT} systems,'' in
  \emph{Annual American Control Conference (ACC 2018)}, Milwaukee, WI, USA,
  Jun. 2018.

\bibitem{spatiotemporal_estimation}
A.~Malik, R.~Maciejewski, S.~Towers, S.~McCullough, and D.~S. Ebert,
  ``Proactive spatiotemporal resource allocation and predictive visual
  analytics for community policing and law enforcement,'' \emph{IEEE
  Transactions on Visualization and Computer Graphics}, vol.~20, no.~12, pp.
  1863--1872, Dec. 2014.

\end{thebibliography}

\end{document}